\documentclass{ptapap}

\usepackage{array}
\usepackage{url}
\usepackage{natbib}
\usepackage{amssymb}	% Extra maths symbols

\def\ms{$\mathfrak{M}_{\odot}$}
\def\oc{{\it O-C}}

\author{Marek Skarka}[ASU,KONK,VSES]
\author{Ji\v{r}\'{i} Li\v{s}ka}[VSES,CEIT]
\author{\'{A}dam S\'{o}dor}[KONK]
\author{Elisabeth Guggenberger}[GOT,AAR]
\author{Radek D\v{r}ev\v{e}n\'{y}}[VSES,ZNOJ]
\author{Thomas Barnes}[MCD]
\author{Katrien Kolenberg}[LEUV,ANTW]

\affil[ASU]{Astronomical Institue, Czech Academy of Sciences, Fri\v{c}ova 298, 251 65 Ond\v{r}ejov, Czech Republic}
\affil[KONK]{Konkoly Observatory, Research Centre for Astronomy and Earth Sciences, Hungarian Academy of Sciences, Konkoly Thege Miklós út 15-17, H-1121 Budapest, Hungary}
\affil[CEIT]{Central European Institute of Technology - Brno University of Technology (CEITEC BUT), Purky\v{n}ova 656/123, CZ-612 00 Brno, Czech Republic}
\affil[GOT]{Max Planck Institut f\H{u}r Sonnensystemforschung, Justus-von-Liebig-Weg 3, D-37077 G\"{o}ttingen, Germany}
\affil[AAR]{Stellar Astrophysics Centre, Department of Physics and Astronomy, Aarhus University, DK-8000 Aarhus C, Denmark}
\affil[VSES]{Variable Star and Exoplanet Section of the Czech Astronomical Society, Vset\'{i}nsk\'{a}~78, Vala\v{s}sk\'{e} Mezi\v{r}\'{i}\v{c}\'{i}, CZ-757 01, Czech Republic}
\affil[ZNOJ]{Znojmo Observatory, Vinohrady 57, CZ-669 02 Znojmo, Czech Republic}
\affil[MCD]{The University of Texas at Austin, McDonald Observatory, 1 University Station, C1402, Austin, TX 78712, USA}
\affil[LEUV]{Institute of Astronomy, KU Leuven, Celestijnenlaan 200D, B-3001 Heverlee, Belgium}
\affil[ANTW]{Physics Department, University of Antwerp, Groenenborgerlaan 171, B-2020 Antwerpen, Belgium}

\title{On the interpretation of the long-term cyclic period variations in RR Lyrae stars}
\headtitle{Long-term cyclic period variations in RR Lyrae stars}

\begin{document}

\maketitle

\begin{abstract}

Many RR Lyrae stars show long-term variations of their pulsation period, some of them in a cyclic way. Such behaviour can be attributed to the light-travel time effect (LTTE) caused by an unseen
companion. Solutions of the LTTE often suggest very eccentric orbits and minimal mass of the companion on the order of several solar masses, thus, in the black hole range. We discuss the
possibility of the occurrence of the RR Lyr-black hole pairs and on the case of Z~CVn demonstrate that the LTTE hypothesis can be false in some of the binary candidates.

\end{abstract}

\section{Introduction}\label{Sect:Introduction}
It is generally accepted that large portion of the stars in the Universe is bound in binary and multiple systems. For example, \citet{raghavan2010} found that among A-type main-sequence stars at least 70\,\% of them have a companion. If we consider the progenitor of an RR Lyrae (RRL) type star as a low-mass star of G-type, we still should observe about 40-50\,\% binaries. Of course, RRL stars are old stars that already passed the red-giant evolutionary phase, which would definitely affect the incidence rate of binaries with RRL stars. 

However, depending on the mass (0.1-1.3 \ms), the percentage of population II stars in binaries ranges from about 20\,\% to 40\,\% \citep{duchene2013}. The orbital period of G-type main sequence stars peaks at $\log(P)$(days)$=4$ suggesting large number of stars with very wide orbits preventing mass transfer \citep{duchene2013}. Although there is no available RRL binary statistics, we could basically expect that many of RRL stars untouched by the evolution of a close companion would be found in binary systems.

This is in clear contrast with what we observe. There are about 100\,000 RRL known in our Galaxy, LMC and SMC \citep{watson2006,soszynski2014,soszynski2016}, but only 82 candidates for binary systems \citep{liska2016c}\footnote{See the online version of the list at \url{http://rrlyrbincan.physics.muni.cz/}.} and only one (almost) confirmed binary \citep[TU UMa,][]{wade1999,liska2016a}. Thus, the binarity of RRL stars remains puzzling, which leaves us with the RRL mass estimated only roughly from the evolutionary and pulsation models. 

\section{The Light-travel time effect and minimal masses for the companions}\label{Sect:LTTE}
According to discussion in \citet{skarka2016a} it is hard to discover an RRL star in a binary using direct methods (eclipses, colour excess, radial velocity measurements) due to large luminosity of RRL stars. Thus, the most promising, and so far the most effective, method is the investigation of cyclic period variations that are considered to be a result of the orbital motion and finite speed of light -- the well known Light-Travel Time Effect, LTTE. 

A big advantage of this method is that also low-mass companions can be detected and by modelling the period variations (usually in the form of {\oc~ diagrams}\footnote{{\it `O'} stands for observed and {\it `C'} stands for calculated.}) we can get orbital parameters, orientation of the orbit in space, and mass function. Thus, investigation of cyclic period variations of binary systems is a full alternative to radial velocity observations. On the other hand, it is very time demanding and only binaries with orbital periods longer than hundreds of days can be investigated.

Based on the investigation of LTTE, 52 stars have been discovered \citep[see the online list by and references therein][]{liska2016c} and additional candidates have been announced at this conference \citep[see][this proceedings]{hajdu2017,greer2017,poretti2017}. The function of mass or minimal mass was estimated in a few studies \citep[for example,][]{hajdu2015,liska2016b}. A very disturbing fact is that in some of the candidates the calculated minimal mass is in the range of black holes \citep{liska2016b,sodor2017} in some cases with masses significantly above 10\,\ms \citep{derekas2004}. Our analysis of the \oc~in Z~CVn suggests even larger minimal mass.

\section{Z CVn}\label{Sect:ZCVn}

In \citet{skarka2017}, we performed an extensive study of Z CVn employing various methods and approaches: from photometry (53 nights during four observing seasons) and spectroscopy (eleven nights during two seasons) to LTTE analysis (227 maximum times, 72 new). We also used archival data, when available. Besides the deep analysis of the Blazhko effect and physical characteristics, the corner stone of our study was the investigation of period variations.

\begin{figure}
\includegraphics[width=\textwidth]{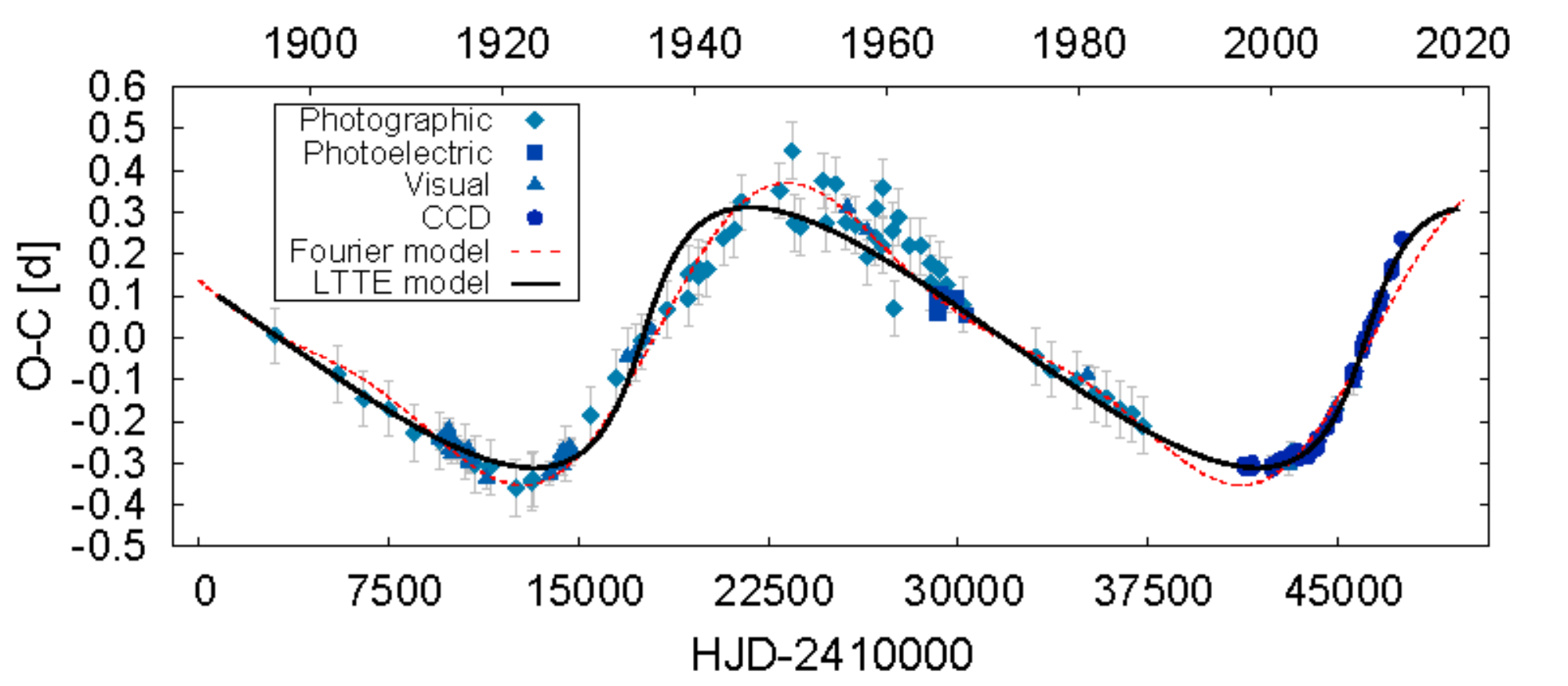}
\caption{The \oc~diagram of Z CVn consisting of our measurements and archival times of maximum light. The meaning of different symbols and lines is clear from the legend.}
\label{Fig:ZCVn}
\end{figure}

The spectacular cyclic period variation is apparent from the \oc~diagram shown in Fig. \ref{Fig:ZCVn}. The data gathered using various methods (shown with different symbols) are complemented with the model when the LTTE is considered (black continuous line) and Fourier fit with two harmonics with period of 78 years (for comparison).

The full set of orbital parameters that we obtained can be found in \citet{skarka2017}. The most important for us now is the minimal mass that was estimated as 56.5\,\ms, which is very suspicious. 

\section{Discussion}\label{Sect:Discussion}
First we asked a question whether such massive black hole could exist. Due to the discovery of the merging black holes detected by the LIGO experiment, when two black holes with masses of 29 and 36\,\ms~coalesced into one object \citep{abbott2016}, we now know that such objects can exist. Also from the theory of formation of black holes it is clear that massive black holes can come to existence from very massive stars that have strong magnetic fields and low metallicity causing reduction of the stellar wind \citep{spera2015,petit2017}.

The second question to ask is whether we can detect some (X-ray) radiation from the material falling onto the black hole. The classical mass transfer known from close binary systems can be ruled out in such wide system. Thus, the only source of material could be the stellar wind from the RRL component and the surrounding interstellar matter. The accretion would take place via the Bondi-Hoyle-Lyttleton accretion \citep{hoyle1939,bondi1944,bondi1952} which is inversely proportional to $\left( v_{\rm s}^{2}+v_{\rm r}^{2}\right)^{3/2}$, where $v_{\rm s}$ is the speed of sound in the material and $v_{\rm r}$ is the relative speed of the black hole with respect to the surrounding material.

Unfortunately, we do not know the density of the stellar wind from RRL stars and temperature (both define $v_{\rm s}$) of the surrounding material. Thus, the resulting luminosity of the black hole could be orders of magnitude lower, but also orders of magnitude larger than the luminosity of Z CVn itself (depending on the input density, temperature and relative speed). In any case, there is no X-ray source detected in the direction to Z CVn.

Another chance to detect a black hole is via the supernova remnant. The problem is that the remnants survive maximally for tens of thousands of years. If the progenitor of the black hole exploded, it necessarily had to happen billions of years ago. In addition, we do not know whether Z CVn corotates with the Galactic halo and, thus, whether the possible remnant follows Z CVn. Theoretical studies show that massive stars may not explode as a supernova, but collapse directly to a black hole \citep{fryer1999}. In any case, there are no signs of the remnant. 

\begin{figure}
\includegraphics[width=\textwidth]{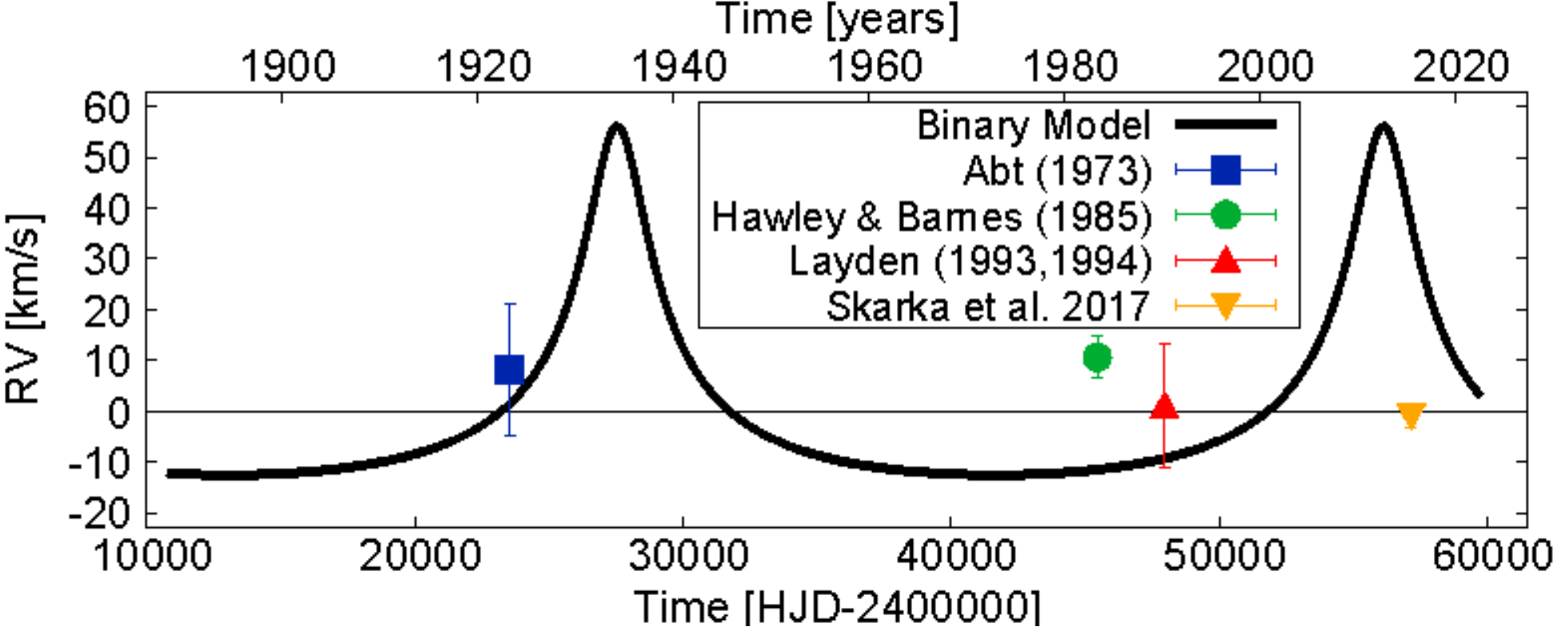}
\caption{RV measurements (points) versus RV model based on LTTE assumption.}
\label{Fig:RV}
\end{figure}

However, the most important evidence against binarity of Z CVn are the spectroscopic measurements. We complemented our RV measurements \citep[11,][]{skarka2017} with available observations from literature \citep{abt1973,hawley1985,layden1993,layden1994} and got the systemic radial velocities (Fig. \ref{Fig:RV}). It is seen that the measurements do not match the model based on LTTE at all, which, without any doubt, rejects the binary hypothesis.

\section{Conclusions}\label{Sect:Conclusions}
In \citet{skarka2017} we showed that although massive black holes can exist, the period variations of Z CVn cannot be caused by an unseen companion implying existence of some unknown long-term effect that is intrinsic to the star.

\acknowledgements{The financial support of the Hungarian NKFIH Grants K-115709, K-113117 and K-119517 (MS, \'{A}S), Czech Grant GA \v{C}R 17-01752J (MS), project CEITEC 2020 (LQ1601) (JL) are acknowledged.}

%\acknowledgements{The financial support of the Hungarian NKFIH Grants K-115709, K-113117 and K-119517 is acknowledged. MS was additionally supported by the Czech Grant GA \v{C}R 17-01752J and acknowledges the support of the postdoctoral fellowship programme of the Hungarian Academy of Sciences at the Konkoly Observatory as host institution. \'{A}S was supported by the J\'{a}nos Bolyai Research Scholarship of the Hungarian Academy of Sciences. This research was carried out under the project CEITEC 2020 (LQ1601) with financial support from the Ministry of Education, Youth and Sports of the Czech Republic under the National Sustainability Programme II (JL).}

%\bibliographystyle{ptapap}
%\bibliography{ptapapdoc}

\end{document}